\documentclass[aps,final,showpacs,showkeys,twocolumn,nofootinbib,superscriptaddress]{revtex4}
\usepackage{graphicx,epsfig}
\usepackage{amssymb}
\usepackage{amsmath}
\usepackage{showkeys}

\def\be{\begin{equation}}
\def\ee{\end{equation}}
\def\ba{\begin{eqnarray}}
\def\ea{\end{eqnarray}}

\def\c2{{c_s^2}}

\def\cp{{x_*^+}}
\def\cm{{x_*^-}}
\def\ai{{a_\infty}}
\def\aisq{{a_\infty^2}}
\def\pd{{\partial}}
\def\L{{\mathcal{L}}}
\newcommand{\bse}{\begin{subequations}}
\newcommand{\ese}{\end{subequations}}

\begin{document}

\title{Perfect fluid and scalar field in the Reissner-Nordstr\"om metric}

\author{\firstname{E.~O.}~\surname{Babichev}}
\email{eugeny.babichev@physik.uni-muenchen.de}
\affiliation{Arnold-Sommerfeld-Center for Theoretical Physics, \\
Department f\"ur Physik, Ludwig-Maximilians-Universit\"at M\"unchen, \\
Theresienstr. 37, D-80333, Munich, Germany} \affiliation{Institute
for Nuclear Research of the Russian Academy of Sciences, 60th
October Anniversary Prospect, 7a, 117312 Moscow, Russia}
\author{\firstname{V.~I.}~\surname{Dokuchaev}}
\email{dokuchaev@ms2.inr.ac.ru} \affiliation{Institute for Nuclear
Research of the Russian Academy of Sciences, 60th October
Anniversary Prospect, 7a, 117312 Moscow, Russia}
\author{\firstname{Yu.~N.}~\surname{Eroshenko}}
\email{eroshenko@ms2.inr.ac.ru} \affiliation{Institute for Nuclear
Research of the Russian Academy of Sciences, 60th October
Anniversary Prospect, 7a, 117312 Moscow, Russia}


\begin{abstract}
We describe spherically symmetric steady-state accretion of perfect
fluid in the Reissner-Nordstr\"om metric. We present analytic
solutions for accretion of a fluid with the linear equations of
state and of the Chaplygin gas. It is also shown that, under
reasonable physical conditions,  there is no steady-state accretion
of a perfect fluid onto a Reissner-Nordstr\"om naked singularity.
Instead, a static atmosphere of fluid is formed. We discuss a
possibility of violation of the third law of black hole
thermodynamics for a phantom fluid accretion.
\end{abstract}

\pacs{04.20.Dw, 04.40.Nr, 04.70.Bw, 96.55.+z, 98.35.Jk, 98.62.Js}
\keywords{black holes, singularities and cosmic censorship}

\maketitle

\section{Introduction}
\label{intro}

The problem of matter accretion onto compact objects in Newtonian
gravity was formulated in the self-similar manner by Bondi
\cite{Bondi}. In the framework of General Relativity steady-state
spherical symmetric flow of test gas onto a Schwarzschild black hole
was investigated by Michel \cite{Michel}. Detailed studies of
spherically symmetric accretion of different types of fluids onto
black holes were further undertaken in a number of works
\cite{accretion}, see also a review \cite{Carr:2010wk}.

In this paper, we study perfect fluids and scalar fields in the
Reissner-Nordstr\"om (RN) metric. We describe spherically symmetric
steady-state accretion of a test perfect fluid with a general
equation of state onto a non-rotating charged black hole. We find
analytic solutions for accretion of a perfect fluid with the linear
equation of state and of the Chaplygin gas onto a RN black hole.
When a phantom fluid accretes onto a black hole, the latter loses
its mass. This result is in consistency with the findings of
Ref.~\cite{bde04} on the phantom accretion onto a Schwarzschild
black hole.

We find that under reasonable physical assumptions a perfect fluid
does not accrete onto the RN naked singularity, i.~e., when
$M^2<Q^2$, where $M$ is the mass and $Q$ is the electric charge of
the naked singularity. Namely, steady-state accretion onto a naked
singularity is only possible in two unphysical cases. In the first
case the accreting fluid is superluminal and an additional boundary
condition on the central singularity is specified. In the second
case, the fluid may be stiff or subluminal, but one has to postulate
that the inflow and outflow coexist in the space-time manifold, and
the solution passes somehow through a singular point. We show, that
instead of a steady-state accretion a static atmosphere around a
naked singularity is formed\footnote{A similar result for Kerr naked
singularity was found  in \cite{Bambi09} using numerical methods.}.

We also show that the extreme state of electrically charged black
hole is reached in a finite time due to phantom fluid accretion,
when gravitational back reaction of an accreting fluid is neglected.
We argue, however, that the test fluid approximation may be violated
when a RN black hole or naked singularity is almost extreme. This
implies  that back reaction of the fluid on the background geometry
may prevent transformation of a black hole into a naked singularity,
in accordance with the third law of black hole thermodynamics
\cite{bch73}.

The paper is organized as follows. In Sec.~{\ref{Sec accretion}} we
construct general formalism for steady-state spherically symmetric
accretion of a test perfect fluid in the  RN metric. In
Sec.~\ref{Nonexistence} we give an alternative description of
accretion in terms of a scalar field. In Sec.~\ref{Sec solutions} we
apply the results of the previous sections to particular examples of
perfect fluid, namely, we study accretion of a fluid with the linear
equation of state and accretion of the Chaplygin gas. A static
atmosphere of fluids around a naked singularity is described in
Sec.~\ref{Sec atmosphere}. Approaching of a black hole to the
extreme state by accretion of phantom fluid and a possibility of
violation of the third law of thermodynamics is discussed in
Sec.~\ref{approach}. We conclude in Sec.~\ref{discussion}.

\section{Steady-state accretion}
\label{Sec accretion}

In this section we study spherically symmetric steady-state
accretion of a test perfect fluid with a general equation of state
in the RN metric. Here we closely follow the approach of
Michel~\cite{bde04} for a similar  study performed for accretion of
a gas in the Schwarzschild metric.

The RN metric reads,
\begin{equation}
 \label{RN0}
 ds^2=fdt^2-f^{-1}dr^2-r^2(d\theta^2+\sin^2\!\theta\,d\phi^2),
\end{equation}
where
\begin{equation}
 \label{RN1}
 f=1-\frac{2M}{r}+ \frac{Q^2}{r^2}.\nonumber
\end{equation}
Here $M$ is a black hole (or naked singularity) mass, and $Q$ is its
total charge. It is convenient to introduce dimensionless
coordinates,
\begin{equation}
\tau\equiv \frac{t}{M}, \quad x\equiv \frac{r}{M},\nonumber
\end{equation}
and dimensionless electric charge of the black hole $e\equiv Q/M$.
In the case $e^2<1$ the equation $f(x)=0$ has two roots,
\begin{equation}
 \label{hors}
 x_{\pm}=1\pm \sqrt{1 - e^2}.\nonumber
\end{equation}
The larger root, $x=x_+$, corresponds to the event horizon of the RN
black hole, and $x=x_-$ is the so-called Cauchy (or inner) horizon.
In the opposite case, $e^2>1$, the RN metric (\ref{RN0}) describes a
naked singularity without event horizon. The marginal case $e^2=1$
corresponds to an extreme black hole.

The energy-momentum of a perfect fluid reads,
\begin{equation}
 \label{emtensor}
    T_{\mu\nu}=(\rho+p)u_\mu u_\nu - pg_{\mu\nu},
\end{equation}
where $\rho$ and $p$ are the fluid energy density and pressure
respectively, and $u^\mu=dx^\mu/ds$ is the fluid four-velocity with
normalization condition, $u^{\mu}u_{\mu}=1$. We assume that the
pressure is an arbitrary function of the density alone, $p=p(\rho)$.
To find integrals of motion, we use the projection of the equation
for conservation of the energy-momentum tensor onto the 4-velocity,
$u_{\mu}T^{\mu\nu}_{\quad ;\nu}=0$. This gives the continuity
equation,
\begin{equation}
 \label{eq2}
  u^{\mu}\rho_{,\mu}+(\rho+p)u^{\mu}_{\; ;\mu}=0.
\end{equation}
Integrating (\ref{eq2}) once, we find the following integral of
motion (the energy conservation):
\begin{equation}
 \label{flux}
 ux^2 n=-A,
\end{equation}
where
\begin{equation}
\label{n} n\equiv
\exp\left[\;\,\int\limits_{\rho_{\infty}}^{\rho}\!\!
\frac{d\rho'}{\rho'+p(\rho')}\right], \nonumber
\end{equation}
$u=dr/ds<0$ in the case of inflow motion (accretion), and  $A>0$ is
the constant of  integration, which is related to the radial energy
flux.

Integration of the time component of the conservation law,
$T^{\mu\nu}_{\;\;\; ;\nu}=0$, gives another integral of motion (the
relativistic Bernoulli equation):
\begin{equation}
 \label{eq1}
 (\rho+p)(f+u^2)^{1/2}x^2u=C_1,
\end{equation}
where $u\equiv dr/ds$ and $C_1$ is the constant of integration. From
(\ref{flux}) and (\ref{eq1}) one can easily obtain:
\begin{equation}
\frac{(\rho+p)}{n}(f+u^2)^{1/2}=C_2,
 \label{energy}
\end{equation}
where
\begin{equation}
\label{C2} C_2\equiv
\frac{-C_1}{A}=\frac{\rho_\infty+p(\rho_\infty)}{n(\rho_\infty)},\nonumber
\end{equation}
here $\rho_\infty$ is the energy density at infinity. Equations
(\ref{flux}) and (\ref{energy}) along with the equation of state
$p=p(\rho)$ form a closed system for accretion onto a RN black hole
(or naked singularity). This system is to be supplied by the
appropriate boundary conditions. The obtained system of  equations
describes accretion of a perfect fluid with a general equation of
state $p=p(\rho)$, and may be applied, in particular, to accretion
of the Chaplygin gas \cite{Chaplygin} or dark energy described by
the generalized linear equation of state \cite{bde04a}.

The constant $C_2$ is fixed by the boundary condition at infinity.
Fixing of $A$ in (\ref{flux}) and, respectively, the flux is more
tricky. It is provided by a physical requirement to have a smooth
transition through the critical sound point (see details, e.~g. in
\cite{Michel}). The resulting solution should be continuous from
infinity down to the black hole horizon. Following \cite{Michel}, we
find relations at the critical point,
\begin{equation}
 \label{cpoint}
  u_*^2=\frac{x_*-e^2}{2x_*^2}, \quad
  c_s^2(\rho_*)=\frac{x_*-e^2}{2x_*^2-3x_*+e^2},
\end{equation}
where $c_s(\rho)\equiv (\partial p/\partial\rho)^{1/2}$ is the sound
speed, and the subscript '$*$' indicates that the values are taken
at the critical point. From (\ref{cpoint}) one can find,
\begin{equation}
 \label{xc}
 x_*^{\pm} =  \frac{1+3c_*^2}{4c_*^2}
 \left\{1\pm\left[1-\frac{8c_*^2(1+c_*^2)}{(1
 +3c_*^2)^2}e^2\right]^{1/2}\right\},
\end{equation}
where $c_*\equiv c_{s}(x_*)$. Critical points exist only if
\begin{equation}
 \label{q}
e^2\le
\frac{\left(1+3c_*^2\right)^2}{8c_*^2\left(1+c_*^2\right)}.\nonumber
\end{equation}
It is worthwhile to note that in contrast to the case of a
Schwarzschild black hole, there are formally two different critical
points, corresponding to the  plus and the minus signs in
(\ref{xc}). Note also that for $e\to 0$ we find  $x_{*}^{-}\to0$.

\begin{figure}
\includegraphics[width=0.48\textwidth]{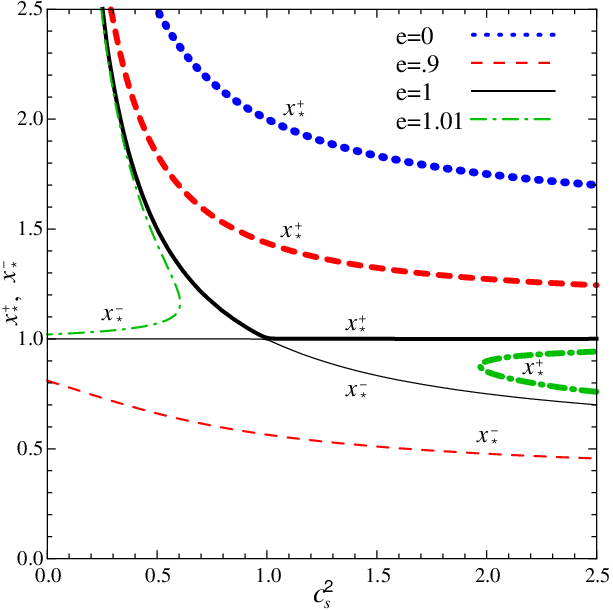}
\caption{The outer critical radius $x_*^+$
(thick lines) and inner critical radius $x_*^-$ (thin lines) are
shown as functions of the sound speed $c_s$ for several values of
the electric charge $e=Q/M$. Note that the outer critical radius
coincides with the event horizon, $x_*^+=1$, for the extreme black
hole ($e=1$) in the case of $c_s\geq1$.} \label{Fig Rcr}
\end{figure}

Depending on the values of $e$ and $c_s$ one can identify the
following five cases:
\begin{itemize}

\item $e< 1$, $c_s^2< 1$ ($c_s^2= 1$). In this case the event and the Cauchy horizons exist, $x^+> x^-$,
as well as both critical points; the outer critical point is outside
the event horizon, $x_*^+> x^+$ ($ x_*^+= x^+$), the inner critical
point is between the event and the Cauchy horizons, $x^- < x_*^- <
x^+$ ($x_*^-=x_-$).

\item $e<1$, $c_s^2> 1$. Similar to the previous case the event and the Cauchy horizons, and both critical points exist; however in this case the outer critical point is in between the event and the Cauchy horizons, $ x^-< x_*^- < x^+$ ($x_*^+=x_-=x_+$); the inner critical point is inside the Cauchy horizon, $ x_*^- < x^+$.

\item $e=1$. The event and the Cauchy horizons coincide, $x^+=x^-=1$ and both critical points exist: \\
    in the subluminal case $\cp>1$ and $\cm=1$; \\
    for a stiff fluid, $\c2=1$, we find $x_*^{\pm} =1$;\\
    in the superluminal case, $\cp=1$ and $\cm<1$;

\item $1<e<3/(2\sqrt{2})$. The RN metric describes a naked singularity (the horizons are absent). Critical points  exist for two different branches, namely, when
 \begin{equation}
\begin{aligned}
\label{subbranch} &&\c2\leq \frac{-4 e^2+3 -4e \sqrt{e^2-1} }{8
e^2-9} \,\,\, {\rm (subluminal)},   \nonumber
\end{aligned}
 \end{equation} or
  \begin{equation}
\begin{aligned}
&&\c2\geq \frac{-4 e^2+3 +4e \sqrt{e^2-1} }{8 e^2-9} \,\,\, {\rm
(superluminal)}. \nonumber
\end{aligned}
 \end{equation}

\item $e\geq 3/(2\sqrt{2})$. The RN metric describes a naked singularity. In contrast to the previous case, the critical points exist only for a subluminal branch
(\ref{subbranch}).
\end{itemize}
In Fig.~\ref{Fig Rcr} the critical radii as functions of the sound
speed are shown for several values of $e$.

Substituting the value of $x_*^{+}$ from (\ref{xc}) into the first
relation in (\ref{cpoint}) and then, in turn, substituting $x_*$ and
$u_*$ expressed in terms of $c_*$ into (\ref{energy}) one finds the
closed equation for $\rho$ at the critical point,
\begin{equation}
\label{master}
 \frac{\rho_*\!+\!p_*}{\rho_\infty\!+\!p_\infty}\frac{n_\infty}{n_*}
 \!=\!\frac{1+3c_*^2+\mathcal{D}}{\sqrt{2\left[1+3c_*^2+4e^2c_*^2(c_*^2-\!1)
 +\mathcal{D}\right]}},
\end{equation}
where
\begin{equation}
\label{D} \mathcal{D} = \sqrt{\left(1+3c_*^2\right)^2 -
8e^2c_*^2(c^2+1)}.\nonumber
\end{equation}
For $e=0$ Eqs.~(\ref{master}) and (\ref{D}) reduces to the equation
for the critical point in the case of the Schwarzschild black hole
\cite{bde04}.

The black hole mass changes at rate $\dot M=-4\pi r^2T_0^{\;r}$ due
to fluid accretion. With the help of (\ref{flux}) and (\ref{energy})
this expression can be written as follows,
\begin{equation}
 \label{evol}
 \dot{M}=4\pi A M^2 [\rho_{\infty}+p_{\infty}].
\end{equation}
From this equation it is clear that accretion of phantom energy,
defined by the condition $\rho_{\infty}+p(\rho_{\infty})<0$, is
always accompanied with decrease of the black hole mass. This is in
accordance with previous findings \cite{bde04}. We would like to
stress that the result is valid for any equation of state
$p=p(\rho)$ with $\rho+p(\rho)<0$.

\section{Perfect fluid as a scalar field}
\label{Nonexistence}

It is well known that the dynamics of relativistic perfect fluid in
the absence of vorticity can be described in terms of a scalar
field. In particular, stiff fluid corresponds to a canonical
massless scalar field. In order to describe more complicated
equations of state one should introduce a generalized non-canonical
scalar-field Lagrangian of the form,
\begin{equation}
 \label{sf action}
 \mathcal{L}=\mathcal{L}(X),\quad
 X\equiv\frac12 \pd_\mu\phi\pd^\mu\phi.
\end{equation}
The energy-momentum tensor corresponding to the Lagrangian (\ref{sf
action}) is \be
T_{\mu\nu}=\mathcal{L}_{X}\nabla_\mu\phi\nabla_\nu\phi -
g_{\mu\nu}\mathcal{L}, \nonumber \ee where the subscript $X$ denotes
the derivative with respect to $X$. The correspondence between
scalar field and a perfect fluid with energy momentum tensor
(\ref{emtensor}) is achieved by the following identifications (see,
e.~g. \cite{bmv07}):
\begin{equation}
 u_{\mu}\equiv\frac{\nabla_{\mu}\phi}{\sqrt{2X}},\nonumber
 \label{u}
\end{equation}
where the pressure $p$ coincides with the Lagrangian density of the
scalar field, $p=\mathcal{L}(X)$, and the energy density is
\begin{equation}
\rho\left(X\right)=2X\mathcal{L}_{,X}-\mathcal{L}.\label{e}\nonumber
\end{equation}
The sound speed can be expressed as
\begin{equation}
 c_s^2=\frac{\mathcal{L}_{,X}}{\rho_{,X}}= \left(1+2X  \frac{\mathcal{L}_{XX}}{\mathcal{L}_X} \right)^{-1}.\nonumber
 \label{sound in kessence}
\end{equation}
Apart from the energy density $\varepsilon$ and pressure $p$ one can
formally define ``particle number density'',
 \be
 n\equiv\exp\left(\int\frac{d\rho}{\rho +p}\right)=\sqrt{X}\mathcal{L}_{,X}.\nonumber
\ee and the enthalpy \be
 h\equiv\frac{\rho+p}{n}=2\sqrt{X}.\nonumber
 \ee
Equations of motion following from (\ref{sf action}) are
\begin{equation}
\label{sf eom}
 \pd_\mu\left( \sqrt{-g}\, \mathcal{L}_X
\,g^{\mu\nu}\pd_\nu\phi \right)=0.
\end{equation}
A steady-state flow is described by the ansatz,
\begin{equation}
\label{ansatz} \phi(t,x)=a_\infty t + \psi(x),
\end{equation}
where the constant $\ai$ defines the ``cosmological'' value of
$\dot\phi$ at spatial infinity. One can easily find that for the
ansatz (\ref{ansatz}),
\begin{equation}
\label{X} X=\frac12\left(\frac{a_\infty^2}{f} -
f\psi'^2\right),\nonumber
\end{equation}
and the equation of motion (\ref{sf eom}) can be integrated once to
give
\begin{equation}
\label{sf eom1}
  x^2 f \mathcal{L}_X  \psi'(x)=\sqrt{2}A.
\end{equation}
Equation (\ref{sf eom1}) is in fact another form of (\ref{eq2}),
written in terms of the scalar field. Moreover, Eq.~(\ref{sf eom1})
is an algebraic equation on function $\psi'$. Thus a general
solution will contain $A$, which should be determined via an analog
of the critical point (\ref{cpoint}). From (\ref{sf eom}), one can
find  $\psi''$ in terms of $\psi'$ (this expression also contains
$\L_X$ and $\L_{XX}$). The critical point is found then by equating
of both the nominator and the denominator of the obtained expression
to zero. As a result, one obtains,
\begin{equation}
\label{cpoint1}
 \psi_*'^2=\aisq \frac{ x_*f_*'}{f^2\left(x_* f_*'+4 f_*\right)},
 \quad f_*\psi_*'^2 \mathcal{L}_{XX}=\mathcal{L}_X.
\end{equation}
which is another form of (\ref{cpoint}). Now, we have three
equations (\ref{sf eom1}), (\ref{cpoint1}) which can be used to find
$\psi_*'$, $x_*$ and $A$. This procedure is fully equivalent to the
fixing of the critical point for the accretion of fluid. This
description is very useful for some particular tasks.

In particular, let us analyze (\ref{sf eom1}) in the limit $x\to 0$.
We have,
\begin{equation}
 \label{X limit}
 2X\sim  \frac{x^2}{e^2}B^2 -\frac{e^2}{x^2}\psi'^2.\nonumber
\end{equation}
Since for the fluid $X>0$, this leads to
\begin{equation}
  \label{Xpsi limit}
   X\to 0, \, \psi'^2\to 0, \quad x\to 0.
\end{equation}
On the other hand, we find from (\ref{sf eom1}),
\begin{equation}
  \label{sf eom limit}
\mathcal{L}_X\psi'\to{\rm const}, \quad x\to 0.
\end{equation}
Combining (\ref{Xpsi limit}) and (\ref{sf eom limit}) we conclude
that a fluid reaches $x=0$ during a steady-state accretion only if
$\mathcal{L}_X\to \infty$ for $X\to0$. This means, in particular,
that a fluid, described by the linear equation of state with $\alpha
\leq 1$, does not reach the central singularity at $x=0$, if
$e\neq0$.

\begin{figure}
\includegraphics[width=0.48\textwidth]{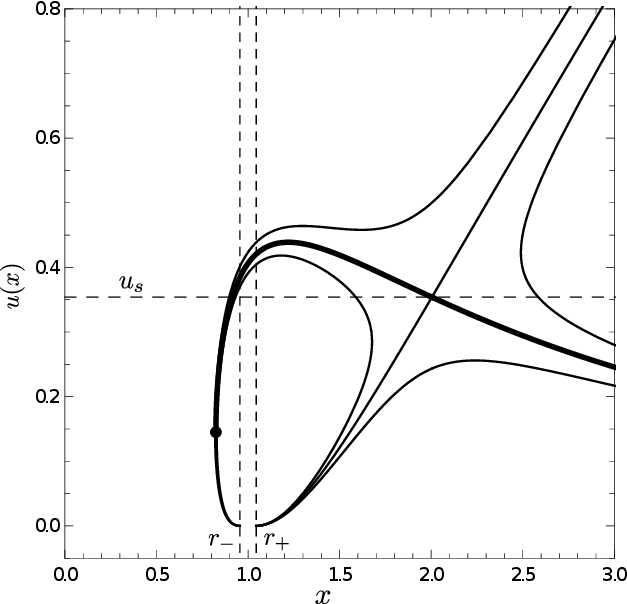}
\caption{Radial 4-velocity $u(r)$ (thick
curve), for the inflowing fluid with $\alpha=1/3$ (thermal photon
gas) in the RN metric with the charge $e=0.999$. Thin curves
correspond to the unphysical hydrodynamical branches and $u_s$ is a
4-velocity at the critical (sound) point.}
 \label{rmine}
\end{figure}

\section{Accretion onto black hole}
\label{Sec solutions}

In this section we present and discuss several analytic solutions
for steady-state accretion of a perfect fluid onto a charged black
hole.

\subsection{Linear equation of state}

As the first example we consider the linear equation of state,
\begin{equation}
\label{leos} p=\alpha(\rho-\rho_0),
\end{equation}
where $\alpha$ and $\rho_0$ are constants. This equation was
introduced in \cite{bde04} (see also \cite{bde04a}) to avoid
hydrodynamical instability for a perfect fluid with the negative
pressure. The constant $\alpha$ in (\ref{leos}) determines square of
the sound speed of small perturbations, $\alpha=c_s^2$, and it must
be positive.  Note, that (\ref{leos}) can be considered as the
linear approximation to a general nonlinear equation of state
$p=p(\rho)$ around some point  $\rho=\rho_1$. Therefore, the results
of this section can be applied to a generic equation of state,
provided that  $|\rho-\rho_1|$  is small enough.

Using (\ref{cpoint}) and (\ref{xc}), one can calculate from
(\ref{flux}) the dimensionless constant $A$ for the linear equation
of state,
\begin{equation}
 A=\alpha^{1/2}x_*^2 \left(\frac{2\alpha x_*^2}{x_*-e^2} \right)^{{\frac{\scriptstyle 1-\alpha}{\scriptstyle 2\alpha}}}.
 \label{Alinear}
\end{equation}
The velocity and the energy density as functions of the radius is
determined by solving (\ref{flux}) and (\ref{energy}),
\begin{equation}
 f+u^2=\!\left(\!-\frac{ux^2}{A}\right)^{\!2\alpha}\!\!, \:
  \frac{\rho\!+\!p}{\rho_\infty+p_\infty}=\!
  \left(\!-\frac{A}{ux^2}\right)^{\!1+\alpha}\!.
  \label{urholinear}
\end{equation}
It is possible to express the solutions of the above equations
through known analytical functions for  specific values of $\alpha$,
namely, $\alpha=1/4$, $1/3$, $1/2$, $2/3$, $1$, $3/2$ and $2$. Below
we present solutions corresponding to some particular values of
$\alpha$.

Let us first consider the case of the stiff fluid: $\alpha =1$. For
the radial velocity and the energy density we find, respectively,
\begin{equation}
\begin{aligned}
 u^2 & = \frac{(x-x_-)x_+^4}{(x+x_+)(x^2+x_+^2)x^2},\nonumber\\
 \rho & = \frac{\rho_0}{2}+\left(\rho_{\infty}-\frac{\rho_0}{2}\right) \frac{(x+x_+)(x^2+x_+^2)}{(x-x_-)x^2}.
 \end{aligned}
\end{equation}
The density at the horizon,
\begin{equation}
\rho_+=\frac{\rho_0}{2}+\left(\rho_{\infty}-\frac{\rho_0}{2}\right)
\frac{2x_+}{\sqrt{1-e^2}}. \label{rhohorizonalpha1}
\end{equation}
Note, that the energy density diverges at the event horizon $x_+$ of
an extreme black hole, $e=1$.

The solutions for thermal photon gas: $\alpha=1/3$ can be found
accordingly. Indeed, radial distribution of the energy density in
this case reads
\begin{equation}
 \rho=\frac{\rho_0}{4}+\left(\rho_{\infty}-\frac{\rho_0}{4}\right)  \left(\frac{1+2z}{3f}\right)^2,\nonumber
 \label{alpha13}
\end{equation}
where
\begin{equation}
 z=\left\{\
\begin{array}{lr}
 \displaystyle{\cos\frac{2\pi-\beta}{3}\,}, &
 x_+\leq x\leq x_*;\\ \\
 \displaystyle{\cos\frac{\beta}{3}}, & x>x_*
 \label{rho}
\nonumber
\end{array}
 \right.
\end{equation}
and
\begin{equation}
 \beta=\arccos\left(1-\frac{27}{2}A^2\frac{f^{\,2}}{x^4}\right)\nonumber
 \label{beta}
\end{equation}
Phantom energy in this particular case corresponds to the choice
$\rho_0>4\rho_\infty$. At the event horizon $x=x_+$ we have,
\begin{equation}
 \rho_+=\rho(x_+)=\frac{\rho_0}{4}+\left(\rho_{\infty}-\frac{\rho_0}{4}\right) \frac{A^2}{x_+^4}.\nonumber
 \label{alpha13b}
\end{equation}

It is also worth to study the case of a superluminal fluid. As an
example we take $\alpha=2$. Now the inflow consists of two
hydrodynamical branches:
\begin{equation}
 u_{1,2}\!=\!\frac{1}{\sqrt{2}}\frac{A^2}{x^4}\sqrt{1\!\pm\!\sqrt{1\!+\! 4f\frac{x^8}{A^4}}}, \;
 \rho_{1,2}\!=\!\left(\!\frac{A}{u_{1,2}x^2}\!\right)^3\!\!.
\label{alpha2}
\end{equation}
At the outer and inner horizons we find
\begin{equation}
 u_{1}(x_{\pm})=\frac{A^2}{x_{\pm}^4}, \quad u_{2}(x_{\pm})=0.\nonumber
\label{alpha2hor}
\end{equation}
The energy density diverges at $r_-$, and the solution does not
exist for $r<r_-$. The behavior of superluminal fluids ($c_s>1$) is
quite unusual. Apart from the transonic solution (\ref{alpha2}),
there is an infinite family of regular at $r>0$ solutions,
parameterized by $A$, with $A>A_*$. These solutions consist of the
only one hydrodynamical branch, and the sonic horizon is absent.
Using a solution with $A>A_*$ one can probe the singularity of a
black hole with small perturbations. In fact, it is not clear how to
choose the ``correct'' physical solution for a superluminal
fluid.\footnote{One can argue, however, that all these problems are
due to the unphysical choice of equation of state (\ref{leos}). Note
that $\rho\to 0$ as $x\to 0$. The equation of state (\ref{leos}) is
unphysical for $\alpha\neq 1$ at $\rho\to 0$, due to the
pathological behavior of equations of motion for $\psi$ in the limit
$\rho\to 0$, as it was shown in \cite{bmv07}. To cure the model
(\ref{leos}) with $\alpha\neq 1$ for small densities, one can modify
equation of state, such that $p \to \rho$ as $\rho\to 0$. For
example, in terms of the scalar filed the following Lagrangian
\begin{equation}
 \label{corrL}
  \L = \left(\sigma+ X\right)^{3/4} -\sigma,
\end{equation}
with $\sigma$ being small, satisfies this requirement, giving also
``superluminal'' fluid $p=2\rho$ for large densities.}

Contrary to accretion of a superluminal fluid, for a subluminal
fluid the solution exists only above some minimal radius $r_{\rm
min}$, $0<r_{\rm min}<r_-$, so that the inflowing fluid does not
reach the central singularity (see Sec.~\ref{Nonexistence}). The
energy density of the fluid has the maximum at $r_{\rm min}$. For
example, $r_{\rm min}=2(\sqrt{2}-1)M$ and $\rho(r_{\rm min})=
(8/3)^2(12\sqrt{2}+17)\rho_\infty$ in the case of accretion of fluid
with $\alpha=1/3$ (thermal photon gas) onto the extremely charged
black hole.

Note that similar behavior was found for geodesic motion of test
particles with a nonzero mass \cite{Carter68,Lopez} in the RN
metric. In particular, the radial component of the 4-velocity for
parabolic radial geodesics (i.~e. for particle with zero velocity at
infinity) is,
\begin{equation}
 \label{parabolic}
 u_p(x)=\pm\frac{\sqrt{2x-e^2}}{x}.
\end{equation}
The particle bounces at $r_{\rm min}=Q^2/(2M)$ and $u_p(r_{\rm
min})=0$ but $|u'_p(r_{\rm min})|=\infty$ according to
(\ref{parabolic}).

The corresponding  solutions for an accreting subluminal fluid are
singular at $r=r_{\rm min}$, namely, $u'(r_{\rm min})=\infty$ and
$\rho'(r_{\rm min})=-\infty$ (although both 4-velocity and the
energy density are finite at $r=r_{\rm min}$). As a result, the
continuity equation (\ref{eq2}) is ill-defined at $r=r_{\rm min}$.
In the following we assume that (i) the fluid can have double-valued
solutions, so that inflow and outflow solutions can coexist in the
same point of the manifold and (ii) it passes through the
singularity in solution at $r=r_{\rm min}$. Formally these
assumptions imply that we can match solutions for inflow and outflow
at $r_{\rm min}$, so that $\rho_{\rm inflow}(x) = \rho_{\rm
outflow}(x)$ and $u_{\rm inflow}(x) = -u_{\rm outflow}(x)$. A
physical interpretation then is as follows: the fluid accretes onto
a black hole, then it bounces at $r_{\rm min}$ and flows outwards to
the asymptotically flat internal spacetime. Since the inflow and the
outflow are symmetric by construction, in the following we will
present the results for the inflow only.

\begin{figure}
\includegraphics[width=0.48\textwidth]{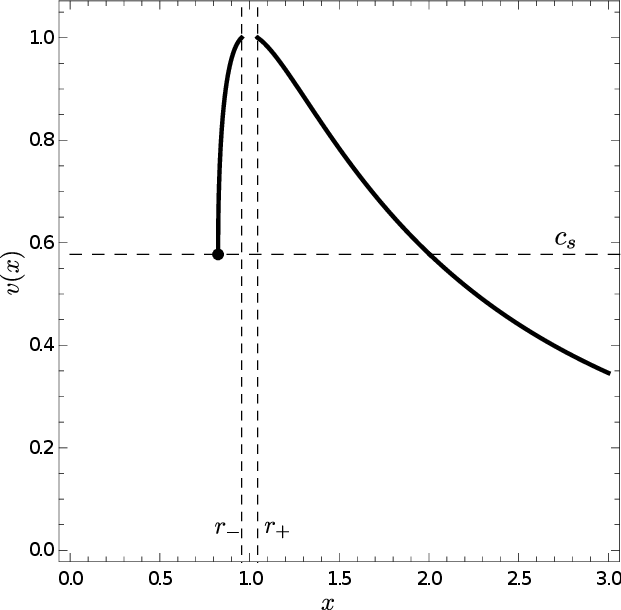}
\caption{Radial 3-velocity $v(x)$ for the
inflowing fluid ($\alpha=1/2$, $e=0.999$) with respect to the local
static observers in the $R$-regions $r_+<r<\infty$ and $0<r<r_-$. In
the $T$-region, $r_-<r<r_+$, the local static observers do not
exist, and thus the 3-velocity is undefined.}
 \label{vqalpha05}
\end{figure}

\begin{figure}
\includegraphics[width=0.48\textwidth]{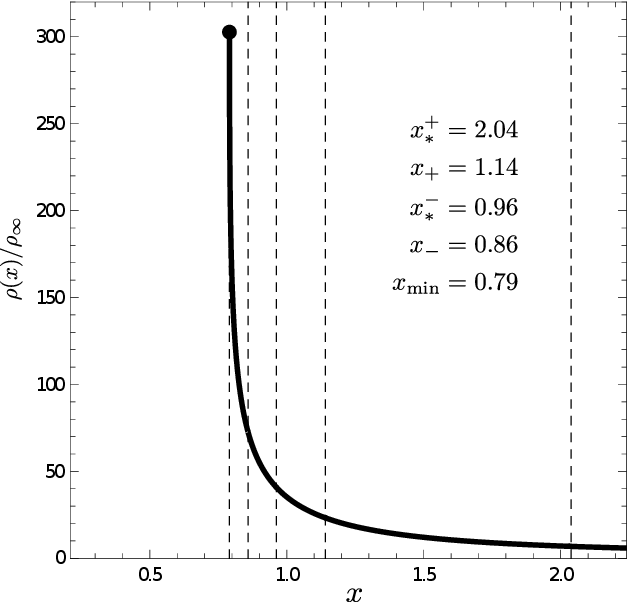}
\caption{Energy density $\rho(x)$, for the
inflowing fluid with $\alpha=1/3$ (thermal photon gas) in the RN
metric with charge $e=0.99$. After reaching the bounce point (marked
by dot) at the minimal radius $r_{\rm min}$, the fluid expands to
the internal asymptotically flat universe.}
 \label{alpha033}
\end{figure}

The resulting distribution for the energy density $\rho(x)$ for the
thermal photon gas is shown in  Fig.~\ref{alpha033}. In
Fig.~\ref{rmine} the corresponding distributions for the radial
component of the 4-velocitiy is shown. In Fig.~\ref{vqalpha05} we
plot the radial 3-velocity $v(x)$ with respect to the local static
observers. Note, that the $v(x)$ equals to the sound speed,
$v(r_{\rm min})=c_s$, at the minimal radius $r_{\rm min}$ for
generic equation of state.

\begin{figure}
\includegraphics[width=0.48\textwidth]{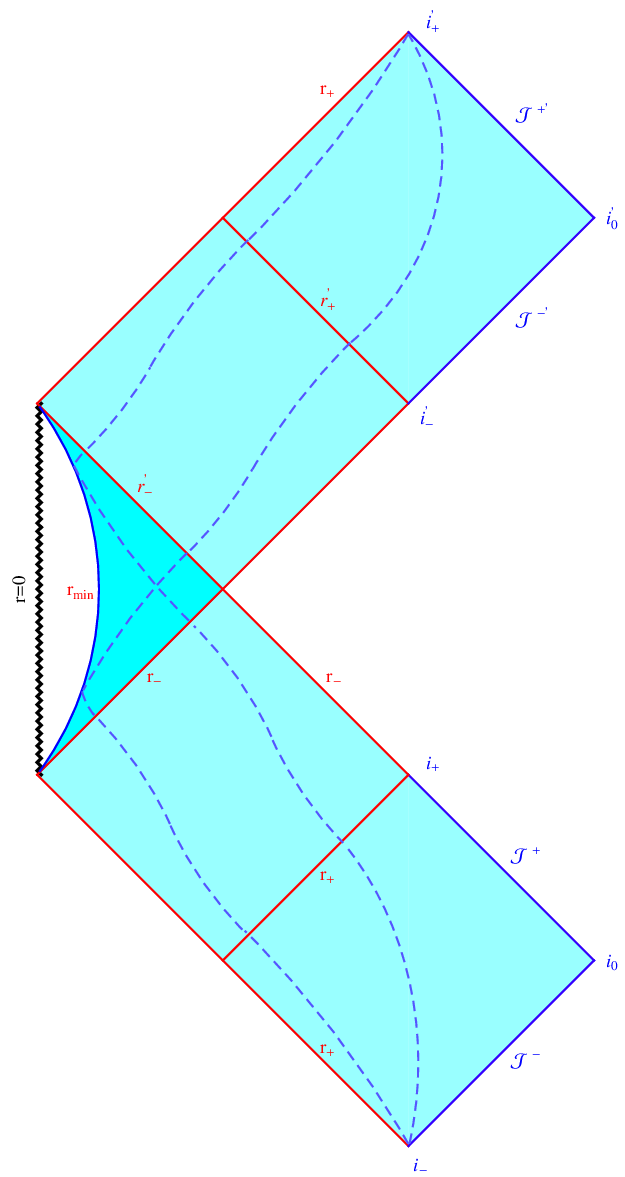}
\caption{Carter-Penrose diagram of the
Reissner-Nordstr\"om metric containing the steady-state accreting
fluid. The streamlines of fluid are shown by the dashed lines. The
minimal radius $r_{\rm min}$ is a bounce point for inflowing fluid.}
 \label{diagram}
\end{figure}

In Fig.~\ref{diagram} we depict a part of the Carter-Penrose diagram
for the the Reissner-Nordstr\"om metric
\cite{GravesBrill60,Carter66}, containing an accreting fluid. This
diagram is symmetric and time-reversal due to the stationarity of
the process. Note that for ``astrophysical'' black holes, formed by
gravitational collapse of massive objects, the internal space-times
are absent and one can expect that inflowing fluid modifies the
metric inside the event horizon (see, e.~g.
\cite{DorNov78,GurSun79,GurNov79,ChanHar82,GneGne93,BanDro95,Burko97,
Brady98,HanKho05,DorHan09,DorHan10} and references herein).

In the Carter-Penrose diagram the streamlines of the outflowing
fluid intersect with the inflowing ones in the region $r_{\rm
min}<r<r_-$ (notice the intersecting dashed lines in
Fig.~\ref{diagram}). As we discussed before, we assume the inflow
and outflow  do not interact and they freely pass through each other
(similar to the motion of test particles). If the fluid is viscous,
the picture should be modified (at least for $r<r_-$, but not for
$r>r_*$), since intersecting streamlines interact. The resulting
flow may become time dependent, turbulent or/and be accompanied by
formation of shocks.

\subsection{Chaplygin gas}

Another analytically solvable example we consider here is the
Chaplygin gas,
\begin{equation}
 \label{chapeos}
 p=-\frac{\alpha}{\rho},\nonumber
\end{equation}
where constant $\alpha>0$ corresponds to a hydrodynamically stable
fluid. The Chaplygin gas with $\rho^2<\alpha$ represents phantom
energy with superluminal speed of sound. The opposite case,
$\rho^2>\alpha$, corresponds to dark energy with $\rho+p>0$ and
$0<c_s^2<1$.

We find the following relations in the critical point:
\begin{equation}
 \label{chapcrit}
 f_*=\frac{\xi\!-\!1}{\xi}, \quad
 x_*^{\pm}=\xi\!\left[1\pm\sqrt{1-\frac{e^2}{\xi}}\right],
 \quad A=\frac{x_*^2}{\sqrt{\xi}},
\end{equation}
where $\xi=\rho_\infty^2/\alpha$.  The sonic point exists and the
accretion is transonic for $\xi\geq e^2$, i.~e. when square root is
real in (\ref{chapcrit}). Note that for the non-phantom Chaplygin
gas this is always satisfied. On the other hand, in the phantom case
the critical point is absent for some range of parameters, implying
that physical solution does not exist. This, however, is merely a
consequence of pathological behavior of Chaplygin gas in the phantom
regime. For radial dependence of the energy density and the radial
4-velocity $u$ we find,
\begin{eqnarray}
 \label{chapdens}
 u&=&-\frac{A}{x^2}\sqrt{\frac{\xi-1}{\xi(\rho/\rho_\infty)^2-1}}, \\
 \frac{\rho}{\rho_\infty}&=&\sqrt{\frac{f-A^2(\xi-1)x^{-4}}{\xi(f-1)+1}}.
\end{eqnarray}
The value of the energy density at the event horizon is
$\rho(r_+)/\rho_\infty=A/x_+^2$. Solution (\ref{chapdens}) in the
specific case $\xi=1$ corresponds to the vacuum state with
$p=-\rho=-\rho_\infty$ and $u=0$. The energy density of the
non-phantom Chaplygin gas diverges at the inner critical point
$x_{\min}=x_*^-=\xi(1-\sqrt{1-e^2/\xi}) $.

\section{Solutions for naked singularity}
\label{Sec atmosphere}

As it was  discussed in Sec.~\ref{Nonexistence}, only
``superluminal'' fluids reach a naked singularity in steady-state
accretion. More precisely, when formulated in terms of a scalar
field, a well-behaved at $r>0$ solution exists only if the
Lagrangian satisfies the relation $d\mathcal{L}/dX\to \infty$ as
$X\to 0$\footnote{As it was discussed in Sec.~\ref{Nonexistence},
for the fluid to be non-pathological the condition
$d\mathcal{L}/dX\to{\rm const}$ as $X\to 0$ must be true. Therefore,
strictly speaking, a ``non-pathological'' superluminal fluid also
does not reach a naked singularity.}. In this case one can specify
the second boundary condition for accretion at the singularity,
$r=0$.

In the case of a ``subluminal'' fluid the critical solution for
steady-state accretion exists not for all $r$, but only for
$r>r_{\rm min}$. This is in fact similar to the case of RN black
hole, when a fluid is bounced from the singularity, as it was
discussed in Sec.~\ref{Sec solutions}. The radial 4-velocity as a
function of $r$ is similar to the case of RN black hole, plotted in
Fig.~\ref{rmine}. The 3-velocity, though, does not have a gap with
undefined values, in contrast to the case of the black hole. If one
thinks in terms of a superfluid, the solution for the critical flow
can be interpreted as two physical solutions: the inflow and
outflow, matched at the point $r_{\rm min}$. Note though, that in
the case of a black hole, the matching point $r_{\rm min}$ (where
the solution becomes singular) is hidden by the horizon, while in
the case of RN naked singularity, the singular matching point is
reachable by a static observer. One should expect that an
arbitrarily small viscosity of the fluid drastically changes the
solution, since the inflowing and outflowing components of the fluid
interact in the whole space-time. Thus we may conclude that for any
realistic fluid the steady-state accretion does not take place for
the RN singularity.

\subsection{Static fluid atmosphere}
\label{ssec fluid atmosphere}

It is interesting, however, that contrary to the black hole case, a
static solution for naked singularity can be constructed. Such a
solution describes a static light atmosphere with zero influx.
Indeed, assuming $u=0$ from (\ref{energy}) we find a static
distribution of a test perfect fluid around RN naked singularity
\begin{equation}
 \frac{\rho+p}{\rho_\infty+p(\rho_\infty)}
 \exp\left[-\int\limits_{\rho_{\infty}}^{\rho}
 \frac{d\rho'}{\rho'+p(\rho')}\right]=f^{-1/2}.\nonumber
 \label{static}
\end{equation}
In the particular case of the linear equation of state (\ref{leos})
we obtain for static atmosphere
\begin{equation}
 \rho(r)= \frac{\alpha\rho_0}{1+\alpha}+
 \left(\rho_\infty-\frac{\alpha\rho_0}{1+\alpha}\right)
 f^{\,-{\frac{\scriptstyle 1+\alpha}{\scriptstyle 2\alpha}}}.
 \label{staticlinear}
\end{equation}
The energy density of ordinary matter (with $\rho_0=0$ and
$\alpha>0$) approaches zero at the singularity, $\rho\propto
x^{1+1/\alpha}$ as $x\to0$. In the case of phantom, the energy
density is finite at $x=0$, and so phantom fluid ``overcomes'' the
naked singularity repulsiveness.

In the case $e^2>1$ by setting $u=A=0$ in the equation
(\ref{chapdens}) we find a static distribution of the Chaplygin gas
around a naked singularity.

\subsection{Static scalar field atmosphere}

Note, that the solutions for static atmosphere of the fluid,
considered above, in Sec.~\ref{ssec fluid atmosphere}, corresponds
to the following solution in terms of a scalar field,
\begin{equation}
 \frac{\pd\phi}{\pd t} = {\rm const},\; \frac{\pd\phi}{\pd r} = 0.\nonumber
\end{equation}
One can notice, however, that zero energy flux, $T^1_0=-f
\mathcal{L}_X\partial_0\phi\partial_1\phi=0$ is also achieved by
setting $\partial _0\phi=0$. Then the equation of motion becomes
\begin{equation}
\frac{\partial}{\partial r}\left(r^2
\mathcal{L}_{X}f\frac{\partial\phi}{\partial r}\right)=0.
\label{sfstatic}
\end{equation}
We restrict our study to the canonical scalar field,
$\mathcal{L}(X)=X$. The solutions of (\ref{sfstatic}) in the case of
RN black hole and a naked singularity are, respectively,
\begin{equation}
\phi(x)=\frac{\xi_1}{M(x_+-x_-)}\ln\left|\frac{x-x_+}{x-x_-}\right|+\xi_2,\nonumber
\label{sfstsol1}
\end{equation}
\begin{equation}
\phi(x)=\frac{\xi_1}{M\sqrt{e^2-1}}{\rm
\arctan}\left[\frac{x-1}{\sqrt{e^2-1}}\right]+\xi_2,
\label{sfstsol2}
\end{equation}
where $\xi_1$ and $\xi_2$ are constants. Note, that $\phi(1)=0$ in
(\ref{sfstsol2}) for any $e\neq0$, but $\phi(0)$ is not necessarily
zero. The energy density of the scalar field is
$T_0^0=\xi_1^2/(2r^4f)$. In the case of a RN black hole it diverges
at the horizon, while for a naked singularity the energy density is
singular at $r=0$. However this singularity is integrable and the
mass of scalar field atmosphere is finite inside any finite $r$.

\section{Approach to extreme state}
\label{approach}

A black hole can approach the extreme state by capturing particles
with electric charge and/or angular momentum, but an infinite time
is required to reach the extreme state
\cite{bch73,bardeen70,roman88}. This is a manifestation of the third
law of the black hole thermodynamics \cite{bch73}. Note, that during
accretion of neutral phantom energy the electric charge of the RN
black hole is unchanged, $Q={\rm const}$, while the black hole mass
decreases. As a result the black hole approaches to near-extreme
state due to the growing of the ratio $e=Q/M(t)$. In the test fluid
approximation, the black hole reaches the extreme state in finite
time $t=t_{\rm NS}$, defined by the relation $Q=M(t_{\rm NS})$.
Indeed, using (\ref{evol}), the time $t_{\rm NS}$ for a black hole
with initial mass $M=M(0)$ and the electric charge $Q={\rm const}$
may be calculated from the following equation,
\begin{equation}
 \label{timeNS}
 \int_0^{t_{\rm NS}}\dot{m}\,dt=Q-M(0).
\end{equation}
If we neglect the cosmological evolution of $\rho_\infty$, then from
(\ref{evol}), (\ref{Alinear}) and (\ref{timeNS}) for the particular
case of phantom with the stiff equation of state ($c_s=1$) we
obtain,
\begin{equation}
 \label{timeNSrn}
 t_{\rm NS}=\frac{e_0^3-3e_0^2+2-2(1-e_0^2)^{3/2}}{3e_0^4}\,\tau,
\end{equation}
where $e_0=Q/M(0)$ and
$\tau=-\{4\pi[\rho_\infty+p(\rho_\infty)]M(0)\}^{-1}$ is the
characteristic accretion time.

The finiteness of time $t_{\rm E}$ in (\ref{timeNSrn}) implies {\sl
violation} of the third law of black hole thermodynamics in the
considered test fluid approximation.\footnote{Possibility for a
black hole to be transformed into a naked singularity by phantom
accretion was first discussed in \cite{MadGon08}.}

Notice, that in deriving the above result we assumed that the fluid
does not back-react. This assumption, however, may not be valid for
the near-extreme black holes/naked singularities. Indeed, in the
case $\alpha\geq 1$, the energy density of the accreting fluid
diverges at the horizon, as the black hole approaches the extreme
state. This can be seen from (\ref{Alinear}),
(\ref{rhohorizonalpha1}) and (\ref{urholinear}). Similarly,
violation of the test fluid approximation occurs at the radius $r=M$
for static atmosphere around near-extreme naked singularity due to
divergence of the energy density, which can be verified from
Eqs.~(\ref{staticlinear}). It is worth to note, that in the case of
near-extreme Kerr-Newman naked singularity the energy density
diverges at $r=M$ for an atmosphere of a fluid \cite{bcde08a}.

Meanwhile, when $0<\alpha<1$ the energy density of the accreting
fluid remains finite even for the extreme black hole. Nevertheless,
one can argue that the test fluid approximation is violated for the
following reason. The test fluid approximation is valid if the back
reaction of an accreting fluid is small. Consider, however, almost
extreme black hole, so that $|m-e|\ll m$. One can calculate the back
reaction from the perturbed Einstein equations,
\begin{equation}
\label{perturbed} \delta G_{\mu\nu} = 8\pi G T_{\mu\nu},
\end{equation}
where $\delta G_{\mu\nu}$ is the deviation of the Einstein tensor
due to the presence of the accreting fluid with the energy-momentum
tensor $T_{\mu\nu}$. Even if perturbation of the metric calculated
from (\ref{perturbed}) is small, in the limit $M\to Q$ a presence of
the fluid may have drastic effect on the metric. Thus one should
carefully consider the back reaction effects in the case of the near
extreme black holes, even if the accreting fluid has a small
energy-momentum tensor. The back reaction of the accretion flow may
prevent conversion of a black hole into a naked singularity
\footnote{The importance of back reaction was discussed in
\cite{hod08} in the context of absorption of scalar particles with
large angular momentum by a near extreme black hole.}. This
question, however, is beyond the scope of this paper, and we leave
it for future investigation.

\section{Conclusion}
\label{discussion}

In this paper we studied steady-state distribution of a test perfect
fluid  with a general equation of state, $p=p(\rho)$, and a scalar
field in the Reissner-Nordstr\"om metric. Similarly to the case of
steady-state accretion of a perfect fluid onto a Schwarzschild black
hole, the corresponding solution for the accretion exists also in
the case of the RN black hole. On the other hand,  no steady-state
accretion of a perfect fluid exists onto the RN naked singularity,
unless one introduces the double-valued velocity, energy density and
the pressure of a fluid, in order to describe the inflow and the
outflow occurring in the same points of space-time. Instead of
steady-state accretion, a static atmosphere of the fluid is formed
around a naked singularity. For both a black hole and naked
singularity we found analytical solutions to the problem of the
steady state configurations of perfect fluids with an arbitrary
equation of state, $p=p(\rho)$. As particular cases, we studied a
fluid with the linear equation of state, $p=\alpha(\rho-\rho_0)$ and
the Chaplygin gas, $p=\alpha/\rho$. We also found a static
distribution of a scalar field around RN naked singularity.

When the accreting fluid is phantom, $\rho+p<0$, the mass of the RN
black hole decreases. This result is in the agreement with the
previous findings \cite{bde04,accrph}. This poses a question,
whether it is possible to {\rm convert} a RN black hole into a naked
singularity by accretion of phantom. Under the assumptions we made,
such a conversion is possible, since the accreting phantom decreases
the black hole mass, while the electric charge of the black hole
remains the same. The conversion of a RN black hole into a naked
singularity in the case of accretion of exotic matter with negative
energy density $\rho<0$ was already studied in
\cite{DorHan10,Sch10}. It is interesting to verify the possibility
of similar conversion in the case of phantom fluid with a positive
energy density $\rho>0$ by taking into account back reaction, which,
as we expect, plays an important role in the case of near-extreme
states. We leave this question for future study.

Although the test fluid approximation seems to break down for the
near-extreme state of the black hole/naked singularity, we would
like to stress that for the far-from-the-extreme state of a black
hole (in particular, for the Schwarzschild solution), the parameters
of the perfect fluid and the boundary condition at the infinity can
be tuned so, that the test fluid approximation describes well the
accretion process.

\begin{acknowledgments}
We would like to thank V.~Beskin, V.~Lukash, Ya.~Istomin, A.~Vikman,
K.~Zybin and S.~Chernov for useful discussions. The work of EB was
supported by the EU FP6 Marie Curie Research and Training Network
``UniverseNet'' (MRTN-CT-2006-035863). The work of other coauthors
was supported in part by the Russian Foundation for Basic Research
grant 10-02-00635 and by the grant of the Leading Scientific Schools
3517.2010.2.
\end{acknowledgments}


\end{document}